\documentclass[pre,aps,twocolumn,floats,superscriptaddress]{revtex4}
\usepackage{graphicx}
\usepackage{amsmath}
\usepackage{amssymb}
\usepackage[usenames]{color}
\newcommand{\bnew}{\color{Black}}
\newcommand{\enew}{\color{Black}}

\def\nin{\noindent}
\def\non{\nonumber}

\def\nin{\noindent}
\def\non{\nonumber}
\def\be{\begin{equation}}
\def\ee{\end{equation}}
\def\bea{\begin{eqnarray}}
\def\eea{\end{eqnarray}}
\def\d{{\rm d}}

\def\r{{\bf r}}

\def\OO{{\cal O}}

\def\lp{\left(}
\def\rp{\right)}
\def\lb{\left[}
\def\rb{\right]}
\def\la{\left<}
\def\ra{\right>}

\def\w{\omega}

\def\Re{\mbox{\,Re\,}}
\def\Im{\mbox{\,Im\,}}

\def\D{{\cal D}}

\begin{document}

\title{Surface-to-volume ratio with oscillating gradients}

\author{Dmitry S. Novikov}
\email{dima@alum.mit.edu}
\affiliation{Center for Biomedical Imaging,
Department of Radiology,
New York University School of Medicine, 660 First Avenue, New York, NY 10016, USA}
\author{Valerij G. Kiselev}
\affiliation{University Medical Center Freiburg,
Department of Radiology, 
Medical Physics,
Breisacher Str.\ 60a,
79106 Freiburg, Germany}

\date{February 6, 2011}

\begin{abstract}
 \nin
Restrictions to diffusion result in the dispersion of the bulk diffusion coefficient. 
We derive the exact universal high-frequency behavior of the diffusion coefficient in terms of the surface-to-volume ratio of the restrictions. This frequency dependence can be applied to quantify structure of complex samples with NMR using oscillating field gradients and static-gradient CPMG. We also demonstrate the inter-relations between different equivalent diffusion metrics, and
describe how to calculate the effect of restrictions for arbitrary gradient waveforms.
\end{abstract}

\maketitle

\section{Introduction}

The universal short-time behavior \cite{Mitra92} of the diffusion coefficient
\be \label{Dt}
D(t) \simeq D_0 \lp 1 - {4\over 3d\sqrt{\pi}} \cdot {S\over V} \sqrt{D_0 t}\rp ,
\quad D_0\equiv D|_{t=0}\,,
\ee
allows one to determine the surface-to-volume ratio $S/V$ of restrictions in porous materials \cite{latour1993,mair1999} and in biological tissues \cite{latour-pnas}.
Here $D_{0}$ is the unrestricted diffusion coefficient, and  $d$ is the effective spatial dimensionality, with the factor $1/d$ arising from the orientational average of the restrictions assuming their statistically isotropic distribution \cite{Mitra92}.

While Eq.~(\ref{Dt}) has been instrumental in characterizing restrictions in a variety of media, the direct measurement of $D(t)$ with pulse field gradient (PFG) diffusion-weighted NMR at millisecond time scales is often technically challenging, especially in the {\it in vivo} applications.

A promising way to get into the short-time limit is to apply the oscillating gradient (OG) method \cite{stepisnik1981}, where the diffusion weighting is effectively accumulated over many periods of oscillation. In this way, the time scale for the diffusivity (the oscillation period) can be much shorter than the total acquisition time thus enabling practical measurements. A variant of this technique requires a constant diffusion gradient, where the temporal modulation is achieved by applying periodic radiofrequency pulses of the CPMG type \bnew \cite{stepisnik1981,MGSE,callaghan-stepisnik1996}. \enew

In view of applying the oscillating techniques
\bnew \cite{stepisnik1981,MGSE,callaghan-stepisnik1996}, \enew
an immediate question is, what exactly should one substitute for the diffusion time $t$ in  Eq.~(\ref{Dt})?  As $t\sim 1/\w$, where $\w$ is the gradient oscillation angular frequency, 
the right-hand side of 
%the $\sqrt{t}$-term in 
Eq.~(\ref{Dt}) must transform in the frequency domain to
\[
D_{0}\lp 1 - C_d \,\frac{S}{V}\,\sqrt{\frac{D_0}{\w}} \rp, \quad \w\to\infty\,.
%-C_d \, D_0\,\frac{S}{V}\,\sqrt{\frac{D_0}{\w}} \,,\quad \w\to\infty\,.
\]
Quite remarkably, the prefactor $C_d$ in this expression has never been explicitly derived for the OG case. The existing analytical results are concerned with a finite number of echoes
\bnew
\cite{deSwiet1994,fordham1996,sen-axelrod,axelrod-sen,zielinski2003,zielinski2004}.
\enew
Furthermore, there exists a discrepancy between the numerical values of $C_d$ provided by different groups \cite{stepisnik2001,stepisnik2007,zielinski2004,zielinski2005,xu2010}.

In this work, we find the prefactor $C_d$ {\it exactly} both for the OG and CPMG cases [Eqs.~(\ref{ReDw}) and (\ref{Cd-MGSE}) below] in the limit of a large number of oscillations. This limit is practically applicable for high oscillation frequencies in accord with the requirement of short diffusion time for the validity of Eq.~(\ref{Dt}).
We show that the exact prefactor values for the infinite OG and CPMG trains differ by less than 1\% from each other [Eq.~(\ref{Cd-MGSE})], thereby justifying the view of the CPMG method as being basically equivalent to the OG, and validate the approximate numerical values found in Ref.~\cite{zielinski2005} for the CPMG and in Ref.~\cite{xu2010} for the OG. To derive our result, we utilize the recently established equivalence between the PFG and OG diffusivities using the effective-medium description of diffusion in disordered materials \cite{EMT,membranes}.

Here, as in Ref.~\cite{Mitra92}, we do not take into account the confounding effects of heterogeneous magnetic susceptibility or relaxation. These effects generally make the interpretation of the diffusion-weighted measurements challenging \cite{kennan1991,does1999,kiselev2004,zielinski2004}.
For the shortest times, they are less relevant; in particular, the effect of the surface relaxation at the pore walls can be factored out \cite{zielinski2004}. However, the confounding effects can accumulate over the total acquisition time of many oscillations, significantly modifying the apparent $S/V$ ratio.

%%%%%%%%%%%%%%%%%%%%%%%%%%%%%%%%%%%%%%%%%%%%%%%%%%%%%%%%%%%%%%%%%%%%
\begin{figure*}[t]
\includegraphics[width=5in]{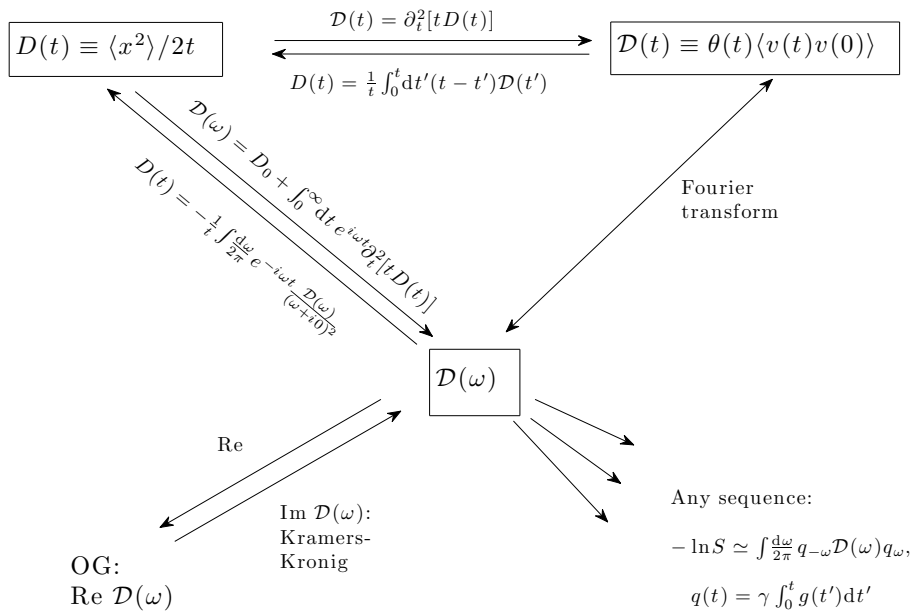}\\
\vspace{-2cm}
\caption{General relations between the three diffusion metrics: $\D(\w)$, $\D(t)$ and $D(t)$, and the signal attenuation.}
\label{fig:Drelations}
\end{figure*}
%%%%%%%%%%%%%%%%%%%%%%%%%%%%%%%%%%%%%%%%%%%%%%%%%%%%%%%%%%%%%%%%%%%%

\section{Methods}

In this Section we outline the relations between the second cumulant of the diffusion-weighted signal and the diffusion characteristics relevant for the PFG and OG measurements, valid for any statistically isotropic disordered medium. In Secton~\ref{sec:results} we will apply these relations to the problem in question.

\subsection{The second cumulant}

We begin with the \bnew Gaussian phase approximation \enew \cite{Callaghanbook,Kiselev2010_diff_book} to the diffusion-weighted signal, $S$,
\be \label{cumt}
-\ln S(T) \simeq \frac12 \int_0^T\! \d t_1 \d t_2\, q(t_1) \la v(t_1)v(t_2)\ra q(t_2) \,,
\ee
which amounts to keeping the second-order term of the cumulant expansion \cite{vanKampen}.
The signal depends on the total duration $T$ of the gradient train $g(t)$, and is a functional of the diffusion-weighting $q(t) = \gamma\int_0^t\! \d t'\, g(t')$, with $\gamma$ the gyromagnetic ratio. The diffusion is characterized by the autocorrelation function $\la v(t_1)v(t_2)\ra$ of molecular velocity, an even function of $t_1-t_2$ in stationary media.
\bnew
As we assumed isotropic diffusion from the outset, $v$ here is the velocity component along the fixed direction of the applied gradient.
\enew
For uniform media,   
$\la v(t_1)v(t_2)\ra = 2D_0\delta(t_1-t_2)$, leading to the standard expression
$-\ln S = bD_0$ with $b=\int_0^T\! q^2(t) \d t$.

Here we will utilize an equivalent, and often more convenient way to represent Eq.~(\ref{cumt}), in terms of Fourier transformed quantities, such as
$q_\w=\int_0^T\! \d t\, e^{i\w t} q(t)$:
\be \label{cum}
-\ln S(T) \simeq \frac12 \int\! {\d\w\over 2\pi}\, q_{-\w} \la v_{-\w}v_{\w}\ra q_\w \,.
\ee
The velocity autocorrelator in the frequency representation is defined as
$\la v_{-\w} v_\w\ra \equiv \int_{-\infty}^\infty \! \d \tau\, e^{i\w\tau} \la v(t_0+\tau)v(t_0)\ra$ independent of $t_0$ due to time translation invariance. The representation (\ref{cum}) underscores that, knowing the correlator $\la v_{-\w}v_\w\ra$, one can evaluate the diffusion-weighted signal (\ref{cumt}) for any gradient waveform $g(t)$. Conversely, by selecting a particular form of $q(t)$ according to its Fourier representation $q_\w$, one effectively allocates a larger or a smaller weight to particular Fourier harmonics $\la v_{-\w}v_\w\ra$ contributing to the measured signal (\ref{cum}).
%\bnew
%Measuring the second cumulant of $S$ is thereby equivalent to a linear filter $|q_\w|^2$ %for $\la v_{-\w}v_\w\ra$.
%\enew

\bnew
There are two advantages of working in the frequency representation (\ref{cum}). From the practical standpoint, a single integral in $\w$ is simpler than a double integral in $t$. This reduction is due to the time translation invariance not explicitly utilized in Eq.~(\ref{cumt}). From the fundamental standpoint, 
$\la v_{-\w}v_\w\ra$ is directly related to the dispersive diffusivity $\D(\w)$ discussed below.
\enew

\bnew
\subsection{The dispersive diffusivity}
\enew

As outlined in detail in Ref.~\cite{EMT}, the dispersive diffusivity $\D(\w)$ is a retarded response function relating the temporal Fourier component ${\bf J}_{\w,\r}=-\D(\w)\nabla_\r \Psi_{\w,\r}$ of the current ${\bf J}(t,\r)$ of diffusing particles to that of a lump of particle density $\Psi(t,\r)$.
\bnew
This makes $\D(\w)$ a central object in the effective medium description of diffusion in disordered media, as it defines the disorder-averaged diffusion equation
\[
-i\w \Psi_{\w,\r} = \D(\w)\nabla_\r^2 \Psi_{\w,\r} + {\cal O}\lp \nabla_\r^4 \Psi_{\w,\r} \rp 
\]
which incorporates the characteristics of the restrictions that can be quantified with a bulk measurement.
\enew
At the same time, $\D(\w) = \int_0^\infty\! \d t\, e^{i\w t} \D(t)$ is the Fourier transform of the retarded velocity autocorrelator
$\D(t)\equiv \theta(t)\la v(t)v(0)\ra$, with $\theta(t)$ a unit step function, cf. Fig.~\ref{fig:Drelations} and Ref.~\cite{EMT}.
Therefore, 
\[
\la v_{-\w}v_{\w}\ra  \equiv 2\Re \D(\w) \,.
\]

\bnew
As a result, the knowledge of $\D(\w)$ allows one to find the second cumulant contribution to the signal attenuation for any pulse sequence $g(t)$ via Eq.~(\ref{cum}):
\be \label{cum-Dw}
-\ln S(T) \simeq \int\! {\d\w\over 2\pi}\, q_{-\w} \D(\w) q_\w \,.
\ee
Here, only $\Re \D(\w)$ contributes, as $\Im \D(\w)$, odd in $\w$, yields zero after being integrated with an even function $|q_\w|^2$. $\Im \D(\w)$ does not contain additional information as it can be restored using the Kramers -- Kronig relations \cite{Landau5}. As we show below, it may be useful to work with the analytic function $\D(\w)$ rather than with its real part.
\enew

The dispersive diffusivity can be obtained  exactly from the narrow-pulse 
PFG diffusion coefficient $D(t)\equiv \la x^2\ra/2t$ via 
\be \label{Dw=Dt}
\D(\w) = D_0 + \int_0^\infty\! \d t\, e^{i\w t} \partial_t^2 \lb t D(t) \rb ,
\ee
where $D_0 \equiv D(t)|_{t=0}$ (cf. Eq.~(D3) in Appendix D of Ref.~\cite{EMT}).
%In the following Section we give the three solutions along these three avenues.
%We underscore that 
The three diffusion metrics: the dispersive diffusivity $\D(\w)$; the retarded velocity autocorrelator $\D(t)$; and the time-dependent diffusion coefficient $D(t)$ contain the same amount of information about restrictions, and thus can be expressed via each other \cite{EMT}, \bnew as illustrated schematically in Fig.~\ref{fig:Drelations}. \enew

\bnew
\subsection{Oscillating gradients}
\enew

A comprehensive diffusion-weighted measurement must provide a way to obtain the diffusivity $\D(\w)$, or the correlator $\la v_{-\w}v_{\w}\ra$, for all $\w$. From this point of view, the OG method, with $g(t)=g_0\cos \w_0 t$, is the easiest one to interpret, as in the limit of the large number
$N= \w_0 T/2\pi\gg 1$ of oscillations,
\[
q_\w = {i\pi \gamma g_0\over \w_0} \lb \delta(\w-\w_0) - \delta(\w+\w_0)\rb
\]
effectively selects the $\w_0$ component $\la v_{-\w_0}v_{\w_0}\ra$, so that
%\be \non
%-\ln S(T)|_{g(t)=g_0\cos\w_0 t} \simeq {(\gamma g_0)^2 \, T\over \w_0^2} \cdot \frac14 \la v_{-\w_0} v_{\w_0}\ra \,.
%\ee
\be \label{OG}
-\ln S(T)|_{g(t)=g_0\cos\w_0 t} \simeq {(\gamma g_0)^2 \, T\over 2\w_0^2} \cdot
\Re \D(\w_0) \,.
\ee
\bnew
Here we used $\delta(\w)|_{\w=0}=T/2\pi$ from the Fourier representation of $\delta(\w)$.
As a result, it is $\Re \D(\w)$ that is measured via the OG techniques \cite{EMT}.
In the above equation, the attenuation over each oscillation period is accumulated, such that the signal $S = \exp\big(-b\cdot \Re \D(\w_0)\big)$ with $b=N b_1$, $b_1 \equiv \pi (\gamma g_0)^2/\w_0^3$.
For the dispersive $\D(\w)$, the $b$-value alone does not define the measurement: the same value, achieved with different oscillation frequencies $\w_0$, will yield different results for $S$. 
%Hence, in what follows we forego the notion of a $b$-value and work directly with $\ln S$.

Remarkably, the signal $S$ is also sensitive to the initial phase $\varphi$ of the oscillation
$g_\varphi(t)=g_0\cos(\w_0 t - \varphi)$, yielding
\be \label{OG-varphi}
-\ln S(T)|_{g_\varphi(t)} \simeq {(\gamma g_0)^2 \, T\over \w_0^2} \cdot
\lb \frac12 \Re \D(\w_0) + \sin^2\varphi \cdot D(T)\rb ,
\ee
where $D(T)\simeq D_\infty \equiv \D(\w)|_{\w\to 0} = D(t)|_{t\to\infty}$ practically is the tortuosity asymptote, since the latter is typically reached over the sufficiently long total measurement time $T$. Physically, the initial phase $\varphi$ leads to the admixture of the PFG attenuation over the time $T$ due to the nonzero value of $q_{\w}|_{\w\to 0} \propto \sin \varphi$, cf. Ref.~\cite{does2003} for $\varphi=\frac\pi2$.
\enew

Equation (\ref{OG}), \bnew as well as its more general counterpart (\ref{OG-varphi}), \enew link the diffusive response function $\D(\w)$ of any medium to the OG attenuation with $N\gg 1$ oscillations.

The above relations 
%between the diffusion metrics and the second cumulant of the signal 
reduce the original problem to finding the diffusivity $\D(\w)$
%velocity correlator $\la v_{-\w}v_{\w}\ra$
for the system in which the PFG diffusion coefficient is given by Eq.~(\ref{Dt}).
It can be done either by solving the problem \cite{Mitra92} in the $\w$-representation, or from a Fourier transform of the retarded velocity autocorrelator $\D(t)$, or directly from
the time-dependent diffusion coefficient $D(t)$ such as the one in Eq.~(\ref{Dt}), measured by ideal narrow-pulse PFG, via Eq.~(\ref{Dw=Dt}).

%%%%%%%%%%%%%%%%%%%%%%%%%%%%%%%%%%%%%%%%%%%%%%%%
\section{Results}
\label{sec:results}

\subsection{Dispersive diffusivity at high frequencies}

Below we find the high frequency limit of $\D(\w)$ corresponding to Eq.~(\ref{Dt}) in three different ways, in order to demonstrate the inter-relations between the above diffusion metrics \bnew (Fig.~\ref{fig:Drelations}): \enew
(i) directly from Eq.~(\ref{Dt}) using Eq.~(\ref{Dw=Dt});
(ii) from the recent result for the diffusivity restricted by membranes \cite{membranes} derived in the frequency representation from the very beginning;
and (iii) from the velocity autocorrelator $\D(t)$ near a flat impermeable wall \cite{astrid2005,astrid2008}.

(i) The most direct way to obtain $\D(\w)$ is to use the \bnew exact relation \enew (\ref{Dw=Dt}). 
\bnew
In general, this relation allows one to find $\D(\w)$ for all $\w$ knowing $D(t)$ for all $t$. However, it can be also utilized to relate {\it each term} of the expansion of $D(t)$ for short or for long $t$ to the corresponding term of the expansion of $\D(\w)$ for high or for low $\w$, respectively. Such expansions in the (fractional) powers of $t$ or $1/t$ usually have finite (or even zero) convergence radius, hence their frequency counterparts should be used within the corresponding bounds. With all that in mind,
\enew
we substitute the second term of Eq.~(\ref{Dt}) into Eq.~(\ref{Dw=Dt}). Reducing the Fourier integral to the Gamma-function by rotating the integration contour $\w t = e^{i\pi/2}u$,  $\int_0^\infty\! \d t \, e^{i\w t} t^{-1/2} = e^{i\pi/4}\w^{-1/2}\Gamma\lp\frac12\rp = e^{i\pi/4}\sqrt{\pi/\w}$,
we find the universal high frequency limit of the dispersive diffusivity
\be \label{Dw}
\D(\w) \simeq D_0\lp 1 -  {e^{i\pi/4}\over d}\,\frac{S}{V}\, \sqrt{D_0\over \w}\rp , \quad \w \to \infty \,,
\ee
which directly leads to our main result (\ref{ReDw}) below.

\bnew
Equation (\ref{Dw}) is the exact universal high-frequency limit of the dispersive diffusivity in the presence of restrictions, valid in the limit in which Eq.~(\ref{Dt}) applies. Further corrections in the inverse powers of $\w$ will contain information about the permeability and curvature of the barriers, as well as the spatial correlations between them \cite{membranes}. 
%The relation (\ref{Dw=Dt}) provides the correspondence between successive terms of asymptotic expansions in the powers of $t$ and $1/\w$.
%To obtain $\D(\w)$ for all $\w$, one needs to know $D(t)$ for all $t$.
\enew

(ii) In recent Ref.~\cite{membranes}, the problem of diffusion restricted by flat permeable membranes was considered in the frequency representation. At high frequencies, this solution is completely equivalent to that of Mitra et al. \cite{Mitra92}, as in the latter work the impermeable pore walls are approximated by locally flat randomly oriented planes as long as the diffusion length is much smaller than the curvature radius of the walls.
Eq.~(\ref{Dw}) then follows from the $\w\to\infty$ limit of $\D(\w)$ found in Ref.~\cite{membranes}, keeping only the $\OO(\w^{-1/2})$ term.

(iii) Finally, in Refs.~\cite{astrid2005,astrid2008} the correction $\delta \D(t)$ to the one-dimensional velocity autocorrelator $\theta(t)\la v(t)v(0)\ra$ in an impermeable box of size $L$ was expressed as the mean
\be  \non
\delta \D(t)
%\la v(t)v(0)\ra
= \frac{1}L \int_0^L \! \hat v(t) G(x_t,x_\epsilon;t) \hat v(0) \,
\d x_0 \d x_\epsilon \d x_t \d x_{t+\epsilon} \,,
\ee
where $\epsilon \to 0$, $G(x_2,x_1;t)$ is the exact diffusion propagator, and the velocity operator $\hat v(t) \equiv [(x_{t+\epsilon}-x_t)/\epsilon]G(x_{t+\epsilon},x_t;\epsilon)
\to -2D_0 \delta'(x_{t+\epsilon}-x_t)$. Indeed,
when integrated with any smooth function $f(x)$, $\hat v$ gives
\bea\non
\lim_{\epsilon\to 0} \,\int_0^L\! \d x\, \d x_0\, {x-x_0\over \epsilon}\, G(x,x_0;\epsilon) f(x) &
\\ \non
 = -2D_0 \int_0^L\! \d x\, \d x_0 \, f(x) \delta'(x-x_0)
 = & D_0\lb f(L)-f(0)\rb
\eea
since $G$ at short times becomes Gaussian, $G \to G_0(x-x_0;\epsilon)\to \delta(x-x_0)$,
with $\frac{x-x_0}{\epsilon}\,G_0 = -2D_0\, \partial_x G_0$. Hence
% which results in
%$\hat v(t) = -2D_0 \delta'(x_{t+\epsilon}-x_t)$.
%Integrating $\hat v(t)$ and $\hat v(0)$ obtain
\bea \non
\delta \D(t)|_{t>0}= - \frac{2}{L} \, D_0^2 G(0,0,t)
= - \frac{S}{V} \, \frac{D_0^2}{\sqrt{\pi D_0 t}}
\eea
%\bea \non
%\la v(t)v(0)\ra &=& 2D_0\delta(t) - \frac{S}{V} \, D_0^2 G(0,0,|t|)
%\\ \non
%&=&  2D_0\delta(t) - \frac{S}{V} \, \frac{D_0^2}{\sqrt{\pi D_0 |t|}}
%\eea
for $L\gg \sqrt{D_0t}$, where $2/L\equiv S/V$.
%for the velocity component orthogonal to the wall positioned at $x=0$.
In the last expression we used the mirror image result for the propagator
$G(x,x_0;t)=G_0(x-x_0;t)+G_0(x+x_0;t)$ near a wall.
Averaging over the orientations in $d$ dimensions leads to
\be
\label{autocorr}
\D(t) = D_0\theta(t) \lp \delta(t) - \frac1{d\sqrt\pi} {S\over V} \,
{\sqrt{D_0}\over \sqrt{t}} \rp .
\ee
Eq.~(\ref{Dw}) is indeed the Fourier transform of this expression (here the first term should be understood as $\lim_{\eta\to +0} \delta(t-\eta)$, as explained in Appendix D of Ref.~\cite{EMT}).
\bnew This derivation completes the ``triangle" of inter-relations in Fig.~\ref{fig:Drelations}. \enew

\subsection{Oscillating gradients}

Taking the real part of Eq.~(\ref{Dw}) we arrive at our main result,
\be \label{ReDw}
\Re \D(\w) \simeq D_0\lp 1 - C_d \frac{S}{V} \sqrt{D_0\over\w}\rp, \quad C_d=\frac1{d\sqrt{2}}
\ee
in $d$ dimensions.
\bnew
Note that, for the phase-shifted OG sequence,
Eq.~(\ref{ReDw}) defines only the first term in Eq.~(\ref{OG-varphi}), whereas the second term, $\propto D(T)$, depends on the particular system geometry over large spatial scales $\sim \sqrt{D(T)T}$. \enew

The above value of $C_d$ contradicts the calculation reported in Refs.~\cite{stepisnik2001,stepisnik2007}, where the corresponding prefactor $1.11\cdot {4\over\pi}\approx 1.41$ (presumably for the three-dimensional case) is about six times greater than our $C_3$.

In Ref.~\cite{xu2010}, the above result for $d=3$ was represented in the form
$4c/(9\sqrt{\pi})\cdot (S/V)\sqrt{D_0 \Delta_{\rm eff}}$, where $\Delta_{\rm eff}=1/(4f) = \pi/(2\w)$ was called the effective diffusion time \cite{parsons2006}, and the correction factor $c\approx 0.73$ was evaluated numerically. The exact value is $c=3/4$ according to Eq.~(\ref{ReDw}), which indicates a 3\% deviation in the numerical approximation found in Ref.~\cite{xu2010}.

\subsection{CPMG in a constant gradient}

A closely related measurement technique is the CPMG train in the presence of a constant gradient \bnew \cite{stepisnik1981,MGSE,callaghan-stepisnik1996}. \enew
Let the interval between successive echoes be $2\tau$, with the rf pulses applied at
$t=\tau, 3\tau, 5\tau, ... $ (cf. notation of Ref.~\cite{zielinski2005}).
This is equivalent to the box-shaped oscillating gradients $g(t)$ alternating between the values $\pm g_0$ with the frequency $\w_0 = 2\pi/4\tau$, with $4\tau$ being the OG period. The Fourier decomposition of this effective square gradient waveform
\be \non
g(t)=\frac{4g_0}\pi \sum_{k=0}^\infty (-)^k  \frac{\cos\w_k t}{2k+1} \,, \quad \w_k = (2k+1)\w_0\,.
\ee
The corresponding $q(t)$ has the Fourier decomposition
\be \non
q_{\w} = {4\gamma g_0\over i\w_0}\sum_{k=0}^\infty {(-)^k\over (2k+1)^2}
\lb \delta(\w+\w_k) - \delta(\w-\w_k)\rb.
\ee
Substituting it into Eq.~(\ref{cum-Dw}),
we find the signal accumulated over a large measurement interval $T=2n \tau$, $n\gg 1$:
\be \non
-\ln S(T) = {8(\gamma g_0)^2 T\over \pi^2\w_0^2} \sum_{k=0}^\infty {1\over (2k+1)^4} \,
\Re \D(\w_k)\,.
\ee

Using the above expression (\ref{ReDw}), find
\be \label{S-MGSE}
-\ln S(T) = {\pi^2 (\gamma g_0)^2 D_0 T\over 12\, \w_0^2}
\lp 1 - \tilde{C}_d {S\over V} \sqrt{D_0\over \w_0} \rp,
\ee
where
\be \label{Cd-MGSE}
\tilde{C}_d = C_d \cdot {s_{9/2}\over s_4}\approx 0.99351277 \, C_d \,.
\ee
Here
$s_\nu = \sum_{k=0}^\infty (2k+1)^{-\nu} = (1-2^{-\nu})\zeta(\nu)$,
where $\zeta(\nu)$ is the Riemann $\zeta$-function.
In particular, $s_4 = \pi^4/96$.

The $1/\sqrt{\w_0}$ term in the exact result (\ref{S-MGSE}) can be written as
$-\tilde C_d \sqrt{2/\pi}\, S\sqrt{D_0\tau}/V \approx -0.186843 \,S\sqrt{D_0\tau}/V$. The numerical prefactor here agrees well with the approximate numerical limit, $-0.19$ (Ref.~\cite{zielinski2005}), of the calculation for the finite number of pulses performed in the time domain \cite{sen-axelrod,axelrod-sen}.

\section{Discussion}

Our approach shows that the exact prefactor $C_d$, Eq.~(\ref{ReDw}) [and its CPMG modification (\ref{Cd-MGSE})], is as universal and independent on the system geometry, as is the corresponding coefficient $4/3d\sqrt{\pi}$ in the original result (\ref{Dt}).
The simplicity and generality of this derivation underscores the utility of the dispersive diffusivity $\D(\w)$. Keeping the diffusivity complex-valued simplifies calculations in many contexts \cite{membranes,EMT}; taking its real part or relating $\D(\w)$ to $D(t)$ is best left for the very last step.

We also note that in general, the concept of the effective diffusion time for the OG protocols \cite{fordham1996,parsons2006} is well defined only as an order-of-magnitude estimate, $t\sim 1/\w$. Indeed, the relations between $\D(\w)$ and $D(t)$ are nonlocal integral relations in time or in frequency \cite{EMT,membranes}, i.e. to determine $\D(\w)$ one needs to know $D(t)$ for all $t$, and vice-versa, to determine $D(t)$ one needs to know $\D(\w)$ for all $\w$. For the particular short-time limit (\ref{Dt}), it was possible to relate the $1/\sqrt{\w}$ term in the expansion of $\D(\w)$ to the corresponding $\sqrt{t}$ term in $D(t)$, which may prompt one to define some effective diffusion time $t_{\rm eff}=\beta/\w$ so that the relative changes in Eqs.~(\ref{Dt}) and (\ref{ReDw}) are the same (this happens for $\beta=9\pi/32$).
However, this exact proportionality relation generally does not hold for all $\w$, i.e. one cannot define some constant $\beta$ such that
$\Re \D(\w)=D(t)|_{t=\beta/\w}$ for all $t$. Instead, one has to use the exact integral relations \cite{EMT} between these quantities.

The effect of restrictions can be calculated for arbitrary gradient waveform using Eqs.~(\ref{cum-Dw}) and (\ref{ReDw}) or their time-domain counterparts. In particular, one can use a gradient waveform defined as a numerical table in magnet's software.

\section{Conclusions}

In this work we used the equivalence between the description of restricted diffusion in time and frequency domains, to find the exact high-frequency behavior for the frequency-dependent diffusivity in disordered media with restrictions, accessible with the oscillating gradient and static-gradient CPMG protocols. Our results will allow one to determine the surface-to-volume ratio of restrictions using the measurement techniques naturally suitable for the shortest time scales. We also demonstrated how the effective medium approach unifies and relates to each other different diffusion metrics, such as the velocity autocorrelator and the time- and frequency-dependent diffusion coefficients.

\section*{Acknowledgments}
This work was motivated by discussions with Junzhong Xu and Lukasz Zielinski.
It is a pleasure to thank them, as well as Jens Jensen, for numerous helpful comments on the manuscript.

%%%%%%%%%%%%%%%%%%%%%%%%%%%%%%%%%%%%%%%%%%%%%%%%%%%%%%%%%%%%%%%%%%%%%

%\bibliographystyle{elsart-num}
%\bibliography{/Users/Kiselev/prog/include/literatureMRI}

\end{document}